\documentclass{PoS}
\usepackage{amsmath}
\usepackage[normalem]{ulem} 

\newcommand{\hc}{\hbox {H.c.}}

\renewcommand{\Re}{\mbox{Re\thinspace}}
\renewcommand{\Im}{\mbox{Im\thinspace}}

\title{Testing the presence of CP violation in the 2HDM}

\ShortTitle{CP violation in the 2HDM}

\author{B. Grzadkowski\\
        Faculty of Physics, University of Warsaw, Pasteura 5, 02-093 Warsaw, Poland\\
        E-mail: \email{bohdan.grzadkowski@fuw.edu.pl}}

\author{O. M. Ogreid\\
        Bergen University College, Postboks 7030, N-5020 Bergen, Norway\\
        E-mail: \email{omo@hib.no}}

\author{\speaker{P. Osland}\\ 
        Department of Physics,
University of Bergen, Postboks 7803, N-5020 Bergen, Norway\\
        E-mail: \email{Per.Osland@ift.uib.no}}

\abstract{We review CP properties of the Two-Higgs-Doublet model.
In particular, we show that spontaneous CP violation occurs in the parameter space on the
border between regions allowing explicit CP violation and those where there is another minimum, deeper than the one corresponding to $v=246~\text{GeV}$. 
We discuss weak-basis invariants which describe CP violation
and express them through measurable quantities like coupling constants and masses.
Also, we discuss how CP violation is constrained by the LHC Higgs data. Finally, we identify 
effective operators that could be adopted to measure CP-invariants.}

\FullConference{Proceedings of the Corfu Summer Institute 2014 "School and Workshops on Elementary Particle Physics and Gravity",\\
		3-21 September 2014\\
		Corfu, Greece }

\begin{document}
\section{Introduction and notation}
\label{sect:introduction}
The Two-Higgs-Doublet model (2HDM) \cite{Gunion:1989we,Branco:2011iw} is of great interest since it provides a simple framework for additional CP violation \cite{Lee:1973iz}, beyond that of the Standard Model, and may thus facilitate baryogenesis \cite{Riotto:1999yt}. On the other hand, the data on the Higgs particle point to Standard-Model (SM) couplings (for a review, see \cite{Ellis:2015dha}), so it is not clear if Nature is making use of this mechanism. We shall here review some quantities which can be explored in order to identify and quantify CP violation in this model, and point out that CP violation is still possible. 

The scalar potential of the 2HDM shall be parametrized in the standard fashion:
\begin{align}
\label{Eq:pot}
V(\Phi_1,\Phi_2) &= -\frac12\left\{m_{11}^2\Phi_1^\dagger\Phi_1
+ m_{22}^2\Phi_2^\dagger\Phi_2 + \left[m_{12}^2 \Phi_1^\dagger \Phi_2
+ \hc\right]\right\} \nonumber \\
& + \frac{\lambda_1}{2}(\Phi_1^\dagger\Phi_1)^2
+ \frac{\lambda_2}{2}(\Phi_2^\dagger\Phi_2)^2
+ \lambda_3(\Phi_1^\dagger\Phi_1)(\Phi_2^\dagger\Phi_2) 
+ \lambda_4(\Phi_1^\dagger\Phi_2)(\Phi_2^\dagger\Phi_1)\nonumber \\
&+ \frac12\left[\lambda_5(\Phi_1^\dagger\Phi_2)^2 + \hc\right]
+\left\{\left[\lambda_6(\Phi_1^\dagger\Phi_1)+\lambda_7
(\Phi_2^\dagger\Phi_2)\right](\Phi_1^\dagger\Phi_2)
+{\rm \hc}\right\} \\
&\equiv Y_{a\bar{b}}\Phi_{\bar{a}}^\dagger\Phi_b+\frac{1}{2}Z_{a\bar{b}c\bar{d}}(\Phi_{\bar{a}}^\dagger\Phi_b)(\Phi_{\bar{c}}^\dagger\Phi_d).
\label{Eq:pot-other}
\end{align}
CP violation may be introduced by taking at least two of the parameters $m_{12}^2$, $\lambda_5$, $\lambda_6$, and $\lambda_7$ complex.
The doublet fields are decomposed as
\begin{equation}
\Phi_j=e^{i\xi_j}\left(
\begin{array}{c}\varphi_j^+\\ (v_j+\eta_j+i\chi_j)/\sqrt{2}
\end{array}\right), \quad
j=1,2.\label{vevs}
\end{equation}
where in a CP-violating theory the physical neutral states can be expressed as linear combinations of the $\eta_1$ and $\eta_2$ on the one hand, with $\eta_3$ on the other:
\begin{equation} \label{Eq:R-def}
\begin{pmatrix}
H_1 \\ H_2 \\ H_3
\end{pmatrix}
=R
\begin{pmatrix}
\eta_1 \\ \eta_2 \\ \eta_3
\end{pmatrix},
\end{equation}
where $\eta_3 \equiv -\sin\beta\chi_1+\cos\beta\chi_2$ with $\tan\beta=v_2/v_1$ and $R$ is a rotation matrix diagonalizing the mass-squared matrix
\begin{equation}
\label{Eq:cal-M}
R{\cal M}^2R^{\rm T}={\cal M}^2_{\rm diag}={\rm diag}(M_1^2,M_2^2,M_3^2),
\end{equation}
and parametrized e.g. in terms of three rotation angles $\alpha_i$ as \cite{Accomando:2006ga}
\begin{equation}     \label{Eq:R-angles}
R
=\begin{pmatrix}
c_1\,c_2 & s_1\,c_2 & s_2 \\
- (c_1\,s_2\,s_3 + s_1\,c_3) 
& c_1\,c_3 - s_1\,s_2\,s_3 & c_2\,s_3 \\
- c_1\,s_2\,c_3 + s_1\,s_3 
& - (c_1\,s_3 + s_1\,s_2\,c_3) & c_2\,c_3
\end{pmatrix}
\end{equation}
with $c_i=\cos\alpha_i$, $s_i=\sin\alpha_i$. 
Note that there is considerable freedom in presenting the potential, e.g.,
one can perform $U(2)$ rotations on the two doublets, without changing the physics.

\section{Spontaneous vs explicit CP violation}

The presence of CP violation can be determined by investigating parameters of the potential, and vacuum expectation values 
(vevs) of the scalar fields. The latter will be parametrized as follows
\begin{equation}
\hat{v}_1=\frac{v_1}{v}e^{i\xi_1},\quad \hat{v}_2=\frac{v_2}{v}e^{i\xi_2},
\end{equation}
with $v=246~\text{GeV}$.
Then, in order to break CP, at least one of following three invariants\footnote{These are invariant under the $U(2)$ transformations mentioned in section~\ref{sect:introduction}.} has to be non-zero \cite{Lavoura:1994fv,Botella:1994cs,Branco:2005em,Gunion:2005ja}:
\begin{subequations} \label{Eq:ImJ}
\begin{align} \label{eq:im_J1}
\Im J_1&=-\frac{2}{v^2}\Im\bigl[\hat{v}_{\bar{a}}^* Y_{a\bar{b}} Z_{b\bar{d}}^{(1)}\hat{v}_d\bigr], \\
\label{eq:im_J2}
\Im J_2&=\frac{4}{v^4}\Im\bigl[\hat{v}_{\bar{b}}^* \hat{v}_{\bar{c}}^* Y_{b\bar{e}} Y_{c\bar{f}} Z_{e\bar{a}f\bar{d}}\hat{v}_a\hat{v}_d\bigr], \\
\Im J_3&=\Im\bigl[\hat{v}_{\bar{b}}^* \hat{v}_{\bar{c}}^* Z_{b\bar{e}}^{(1)} Z_{c\bar{f}}^{(1)}Z_{e\bar{a}f\bar{d}}\hat{v}_a\hat{v}_d\bigr],
 \label{eq:im_J3}
\end{align}
\end{subequations}
where $Z_{a\bar{d}}^{(1)}\equiv\delta_{b\bar{c}}Z_{a\bar{b}c\bar{d}}$. 
In the following we will discuss a model defined by the fact that there exists a weak basis such that
$\lambda_6=\lambda_7=0$, and vevs are real (we shall refer to this model as the ``2HDM5'').
Then it turns out that all three invariants are proportional to $\Im\lambda_5$.

Spontaneous CP violation refers to the situation when a suitable $U(2)$ transformation can bring the potential into a real form, leaving one vev complex.
The following set of invariants can be used to distinguish between spontaneous and explicit CP violation \cite{Davidson:2005cw,Gunion:2005ja}:
\begin{subequations} \label{Eq:I}
\begin{align}
I_{Y3Z}&=\Im \bigl[Z_{a\bar{c}}^{(1)} Z_{e\bar{b}}^{(1)} Z_{b\bar{e}c\bar{d}}Y_{d\bar{a}}\bigr],\\
I_{2Y2Z}&=\Im \bigl[Y_{a\bar{b}}  Y_{c\bar{d}} Z_{b\bar{a}d\bar{f}} Z_{f\bar{c}}^{(1)}\bigr],\\
I_{3Y3Z}&=\Im \bigl[Z_{a\bar{c}b\bar{d}} Z_{c\bar{e}d\bar{g}} Z_{e\bar{h}f\bar{q}} Y_{g\bar{a}} Y_{h\bar{b}} Y_{q\bar{f}}\bigr],\\
I_{6Z}&=\Im \bigl[Z_{a\bar{b}c\bar{d}}Z_{b\bar{f}}^{(1)} Z_{d\bar{h}}^{(1)} Z_{f\bar{a}j\bar{k}}Z_{k\bar{j}m\bar{n}}Z_{n\bar{m}h\bar{c}}\bigr].
\end{align}
\end{subequations}
If all of these vanish, whereas at least one of the $\Im J_i$ is non-zero, then there is spontaneous CP violation.
On the other hand, if at least one of them is non-zero, the CP violation is explicit.

In the 2HDM5 two of these invariants are actually zero. Then the conditions for spontaneous CP-violation can be expressed as follows~\cite{Grzadkowski:2013rza}
\begin{itemize}
\item SCPV1:
\begin{equation}
4\frac{\mu^2}{v^2}\Re\lambda_5-4\left(\frac{\mu^2}{v^2}\right)^2+(\Im\lambda_5)^2=0 \quad
({\rm or~equivalently} \quad
\Im\left[(m_{12}^2)^2\lambda_5^*\right]=0) \label{Eq:scpv1}
\end{equation}
\item SCPV2: 
\begin{equation}
\lambda_1=\lambda_2,\quad \lambda_1=\lambda_3+\lambda_4+\Re\lambda_5-2\frac{\mu^2}{v^2} \quad
({\rm or~equivalently} \quad
\lambda_1=\lambda_2, \;  m_{11}^2=m_{22}^2),
\label{Eq:scpv2}
\end{equation}
\end{itemize}
where the terminology SCPV1 and SCPV2 distinguishes these two cases. Here, we introduce the abbreviation $\mu^2\equiv \Re m_{12}^2 v^2/(2v_1v_2)$.

\begin{figure}[htb]
\centering
\includegraphics[width=\textwidth]{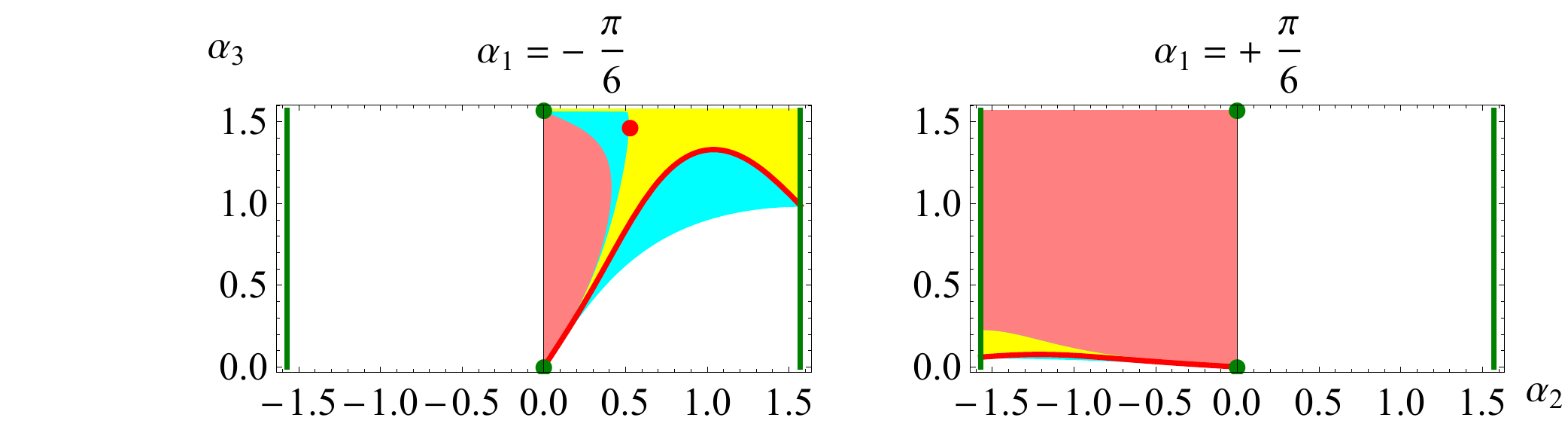}
\includegraphics[width=\textwidth]{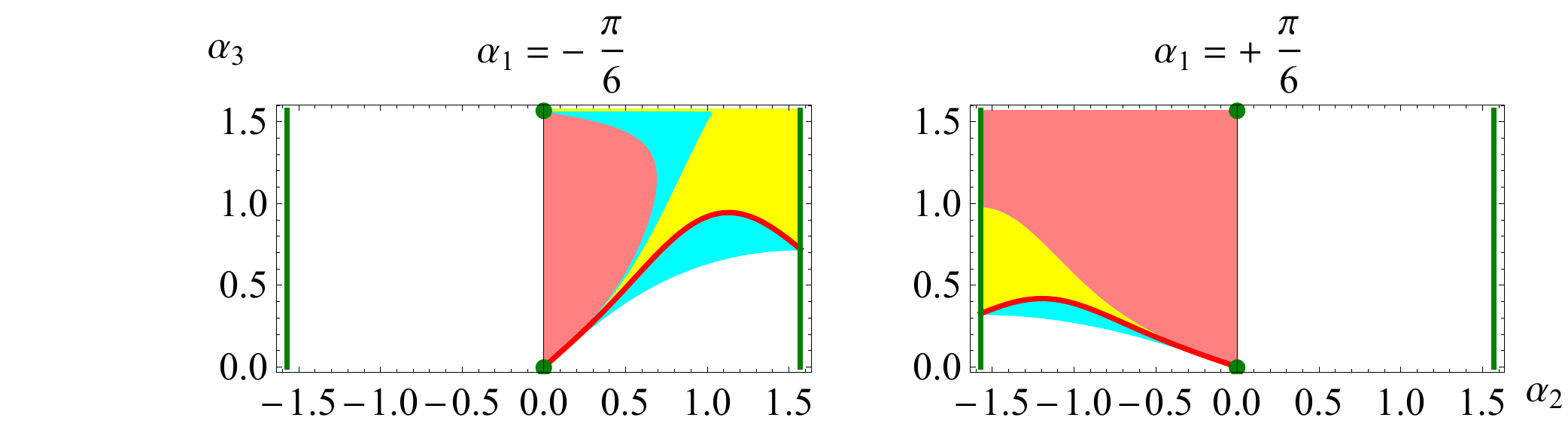}
\caption{Illustration of locations in the $\alpha_2$--$\alpha_3$ space where spontaneous CP violation occurs. Two values of $\tan\beta$ (top: $\tan\beta=2$, bottom: $\tan\beta=5$), and two values of $\alpha_1$ are considered (left: $\alpha_1=-\pi/6$, right: $\alpha_1=+\pi/6$). 
White: mathematically inconsistent; Yellow: explicit CP violation; pink: positivity violated; 
cyan: global-minimum condition violated.
Red curves correspond to parameters that satisfy the SCPV1 condition, {(\protect\ref{Eq:scpv1})}, 
while red dots satisfy the SCPV2 condition, {(\protect\ref{Eq:scpv2})}. 
Both of these indicate spontaneous CP violation.
Green lines and dots indicate locations of CP conservation.}
\label{fig:singlepanel}
\end{figure}

For $M_1=125~\text{GeV}$, $M_2=250~\text{GeV}$, $\mu=300~\text{GeV}$, $M_{H^\pm}=600~\text{GeV}$, and two values of $\tan\beta$, we show in Fig.~\ref{fig:singlepanel} where these conditions are satisfied, and also where CP is conserved. More illustrations of this kind can be found in Ref.~\cite{Grzadkowski:2013rza}.
Note that in this ``simple'' (2HDM5) model, with $\lambda_6=\lambda_7=0$, only one quadrant in the $\alpha_2$--$\alpha_3$ space is accessible, either $\alpha_2<0$ or $\alpha_2>0$ \cite{Khater:2003wq}.

\section{Physical measures of CP violation}
We shall now discuss how to access the quantities (\ref{Eq:ImJ}) experimentally. It has been found that they can be related to the masses of the neutral Higgs bosons, as well as two coupling constants \cite{Lavoura:1994fv,Botella:1994cs,Grzadkowski:2014ada}. In the general case of non-zero $\lambda_6$ and $\lambda_7$, but in a basis with $\xi\equiv\xi_2-\xi_1=0$ (see Eq.~(\ref{vevs})), these required couplings are:
\begin{align}
e_i &\equiv v_1R_{i1}+v_2R_{i2},\label{eq:e_i-def} \\
q_{i}&\equiv\text{Coefficient}(V,H_iH^-H^+)\label{eq:qi}
\nonumber\\
&=\frac{2 e_i}{v^2}M_{H^\pm}^2
-\frac{R_{i2} v_1+R_{i1} v_2}{v_1 v_2}\mu^2
+\frac{g_i}{v^2 v_1 v_2}M_i^2
+\frac{R_{i3} v^3}{2 v_1 v_2}\Im\lambda_5\nonumber\\
&+\frac{v^2 \left(R_{i2} v_1-R_{i1} v_2\right)}{2 v_2^2}\Re\lambda_6
-\frac{v^2 \left(R_{i2} v_1-R_{i1} v_2\right)}{2 v_1^2}\Re\lambda_7.
\end{align}
The quantity $e_i$ describes the coupling of a neutral Higgs $H_i$ to two $Z$ bosons, or to two $W$ bosons. It also describes the coupling of two neutral Higgs particles $H_j$ and $H_k$ to a $Z$ boson, with $i\neq j\neq k\neq i$, and occurs in several other vertices, see \cite{Grzadkowski:2014ada} for details.
The quantity $q_i$ represents the trilinear $H_iH^+H^-$ coupling.
Furthermore, $M_{H^\pm}$ refers to the charged-Higgs mass and we have also introduced the abbreviation
\begin{equation}
g_i\equiv v_1^3R_{i2}+v_2^3R_{i1}. 
\end{equation}

The quantities (\ref{Eq:ImJ}) can then be expressed as \cite{Grzadkowski:2014ada}
\begin{align} \label{Eq:ImJ1}
\Im J_1 
&=\frac{1}{v^5}
\left|
\begin{matrix}
q_1 & q_2 & q_3 \\
e_1 & e_2 & e_3 \\
e_1M_1^2 & e_2M_2^2 & e_3M_3^2
\end{matrix}
\right|, \\[4mm]
\label{Eq:ImJ2}
\Im J_2
&=\frac{2}{v^9}
\left|
\begin{matrix}
e_1 & e_2 & e_3 \\
e_1M_1^2 & e_2M_2^2 & e_3M_3^2 \\
e_1M_1^4 & e_2M_2^4 & e_3M_3^4 
\end{matrix}
\right|, 
\\[4mm]
\Im J_{30} \label{Eq:ImJ3}
&=\frac{1}{v^5}
\left|
\begin{matrix}
e_1 & e_2 & e_3 \\
q_1 & q_2 & q_3 \\
q_1M_1^2 &q_2M_2^2 &q_3M_3^2
\end{matrix}
\right|.
\end{align}
In this form, it is easy to see when any one of them vanishes: two rows or two columns must be proportional.
If any one of the $\Im J_1$, $\Im J_2$ or $\Im J_{30}$ is non-zero, we have CP violation. Whether it is spontaneous or explicit, would be determined by the quantities of Eq.~(\ref{Eq:I}).

\section{CP-violating $ZZZ$ vertex}

\begin{figure}[htb]
\centerline{
\includegraphics[width=5.0cm]{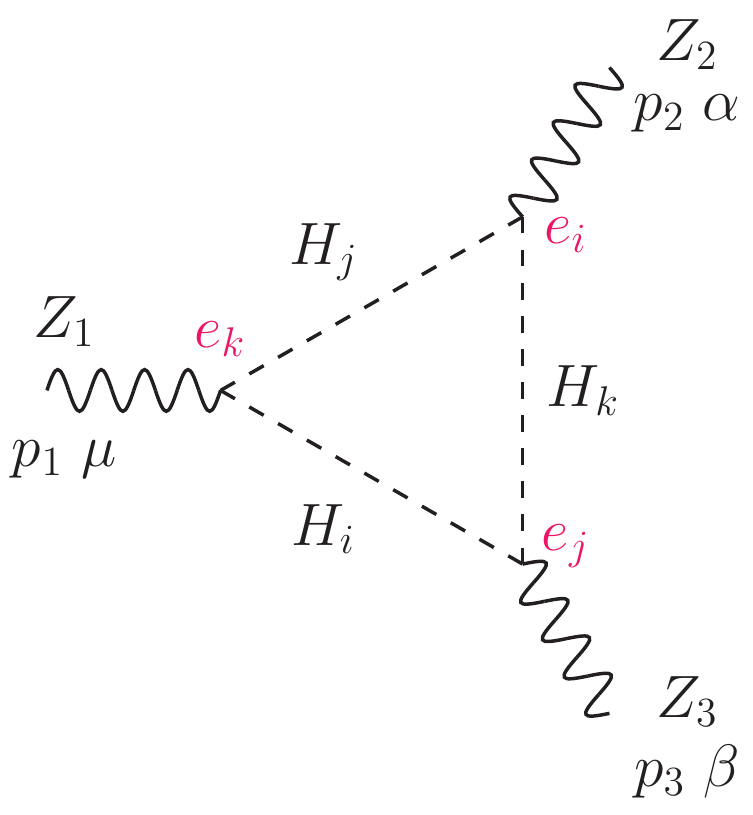}}
\caption{Feynman diagram related to $\Im J_2$.}
\label{fig:Feynman-j2}
\end{figure}

There are various ways in which one could imagine measuring the quantities (\ref{Eq:ImJ1})--(\ref{Eq:ImJ3}) \cite{Grzadkowski:2014ada}. Perhaps the simplest one would be to determine a non-zero and CP-violating contribution to the $ZZZ$ coupling. Another would be the $ZWW$ coupling \cite{GOO}. Such couplings are induced in this model, at the one-loop level, and would be proportional to $\Im J_2$. A characteristic Feynman diagram for the $ZZZ$ case is shown in Fig.~\ref{fig:Feynman-j2}.

In the case when two $Z$ bosons ($Z_2$ and $Z_3$) are on-shell, the conventional parametrization of this vertex \cite{Hagiwara:1986vm,Gounaris:1999kf} is rather simple:
\begin{equation} \label{eq:f_4Z}
e\Gamma_{ZZZ}^{\alpha\beta\mu}
=ie\frac{p_1^2-M_Z^2}{M_Z^2}\left[f_4^Z(p_1^{\alpha}g^{\mu\beta}+p_1^{\beta}g^{\mu\alpha})
+f_5^Z\epsilon^{\mu\alpha\beta\rho}(p_2-p_3)_\rho\right].
\end{equation}
Here, the form factor $f_4$ represents CP violation, whereas $f_5$ represents CP conservation. In the present model, $f_4$ is directly proportional to $\Im J_2$, which again is proportional to $e_1e_2e_3$, whereas $f_5$ gets an SM contribution from fermion loops etc.

Now, $\Im J_2$, and hence this form factor, will depend on the mixing angles of the neutral sector via the product $e_1e_2e_3$. Recent studies constrain the mixing of the CP-even sector (represented by $\eta_1$ and $\eta_2$) with the CP-odd one (represented by $\eta_3$) \cite{Basso:2012st,Basso:2013wna}. Allowed values of $e_1e_2e_3/v^3$ are of the order of ${\cal O}(0.01)$, leading to a prediction for $f_4$ (for masses $M_2$ and $M_{H^\pm}$ of the order of 300--600~GeV) of the order of $10^{-5}$. Current experimental bounds on $|f_4|$ are of the order of $10^{-2}$ \cite{Chatrchyan:2012sga,Aad:2012awa}, so no imminent discovery can be foreseen.

\section{The $H_1$SM (alignment) limit}

The LHC data show that the observed Higgs boson is SM-like.
In particular, the coupling of the observed Higgs particle (which we identify with $H_1$) to vector bosons
is very close to the SM-value, meaning
\begin{equation}
e_1/v=\cos\alpha_2\cos(\alpha_1-\beta)\simeq1,
\end{equation}
or
\begin{equation} \label{Eq:H1SM-limit}
\alpha_1=\beta,\quad \text{and}\quad \alpha_2=0.
\end{equation}

\begin{figure}[htb]
\centerline{
\includegraphics[width=11.0cm]{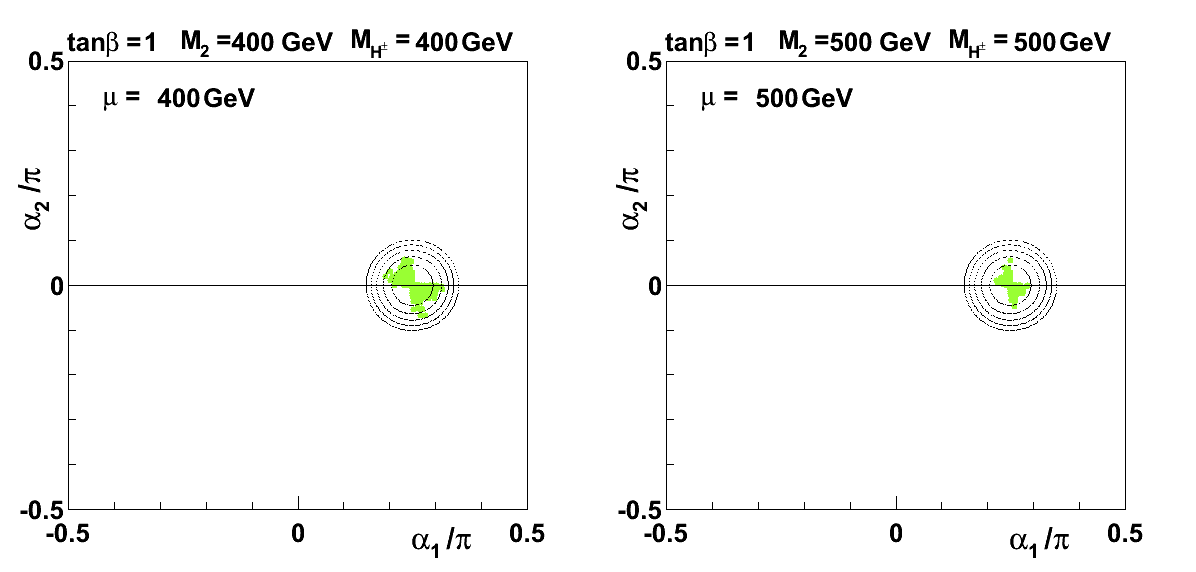}}
\centerline{
\includegraphics[width=11.0cm]{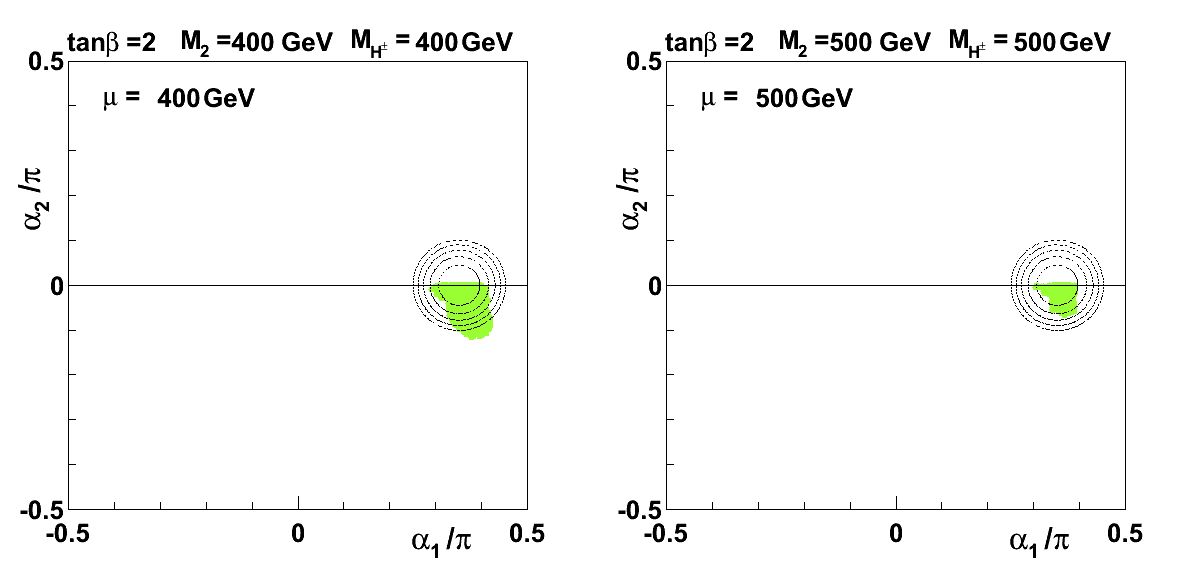}}
\caption{The $H_1$SM (alignment) limit: Allowed regions in the $\alpha_1$--$\alpha_2$ plane for the simplified model with $\lambda_6=\lambda_7=0$.}
\label{fig:H1SM-limit}
\end{figure}

As an illustration, in the simplified model (2HDM5), and for some specific values of $\tan\beta$ and the mass parameters, only the green regions in Fig.~\ref{fig:H1SM-limit} are compatible with LHC data.\footnote{Allowing for $\lambda_6$ and $\lambda_7$ being non-zero, more of this region around the point (\ref{Eq:H1SM-limit}) will be allowed, but it is still confined to small ``radii'' \cite{Grzadkowski:2014ada}.}

Since we must have
\begin{equation}
e_1^2+e_2^2+e_3^2=v^2,
\end{equation}
it follows that
\begin{equation}
e_2\simeq e_3\simeq0.
\end{equation}
We refer to this limit as the $H_1$SM (or alignment) limit.

It is clear from Eqs.~(\ref{Eq:ImJ1})--(\ref{Eq:ImJ2}) that in this case, we have $\Im J_1\simeq\Im J_2\simeq0$. However, there is still the possibility that $\Im J_{30}$ might be non-zero, representing CP-violation in the scalar sector \cite{Grzadkowski:2014ada}. In this limit, (\ref{Eq:ImJ3}) simplifies to
\begin{equation}
\Im J_{30}=\frac{e_1q_2q_3}{v^5}(M_3^2-M_2^2),
\end{equation}
with $e_1=v$.
For this to be non-zero, we must have $q_2\neq0$, $q_3\neq0$, as well as $M_3\neq M_2$.
The first of these conditions means that the two heavier Higgs bosons, $H_2$ and $H_3$, must both have tree-level couplings to $H^+H^-$. Note that this is indeed not possible in the CP-conserving theory,
where the pseudoscalar $A$ does not couple to $H^+H^-$.

In the simplified 2HDM5 model, in general $q_2$ and $q_3$ are both non-zero. However, in that model, when $\alpha_2\to 0$, then also $M_3\to M_2$ \cite{Khater:2003wq,Grzadkowski:2014ada}. This leads to the important conclusion that in order to have CP violation in the $H_1$SM (alignment) limit, we must abolish the 2HDM5 and have
\begin{equation}
\lambda_6\neq0, \quad\text{and/or}\quad \lambda_7\neq0.
\end{equation}
Then, we would have
\begin{equation}
\Im J_1=0, \quad \Im J_2=0, \quad \Im J_{30}\neq0.
\end{equation}

In order to discover CP violation in (or near)  the $H_1$SM (or alignment) limit one 
is forced to look for observables
which are $\propto \Im J_{30}$. As discussed in Ref.~\cite{Grzadkowski:2014ada}, the 
tree-level Feynman diagram for $Z\to H_2H_3\to (H^+H^-)(H^+H^-)$ leads to an amplitude 
proportional to $\Im J_{30}$, as does the loop diagram corresponding to $Z\to H^+H^-$, 
obtained from closing the former tree diagram by connecting an $H^+$ to an $H^-$. 
It would be challenging to access these vertices experimentally.

\section{Summary}

We have reviewed some issues relevant for CP violation within the 2HDM:
\begin{itemize}
\item
We have illustrated how,  in the parameter space, spontaneous CP violation takes place on manifolds that are adjacent to where we have explicit CP violation. At points of spontaneous CP violation, the global minumum of the potential is not unique, but we have two minima of the same depth. In regions of explicit CP violation, the global minimum of the potential is unique, meaning that either there is only one minumum, or there exist two minma of unequal depth.
\item
We review the relationship between invariants defining CP violation and physical coupling constants involving the three neutral Higgs bosons. All three are necessarily involved.
\item
Perhaps the ``simplest'' consequence of CP violation would be the induction of a $ZZZ$ vertex. However, this is very constrained by the LHC Higgs data that show the SM nature of the Higgs boson.
\item
Even in the alignment limit, there could still be CP violation in this model, but then $\lambda_6$ and/or $\lambda_7$ would have to be non-zero. This effect would show up in effective vertices 
where a $Z$ boson is coupled (via neutral Higgs bosons) to two or four charged Higgs bosons.
\end{itemize}
\section*{Acknowledgments}
BG acknowledges partial support by the National Science Centre (Poland) research project, 
decision no DEC-2014/13/B/ST2/03969. PO is supported by the Research Council of Norway.


\end{document}